# GPUs as Storage System Accelerators

Samer Al-Kiswany, Abdullah Gharaibeh, Matei Ripeanu

**Abstract**—Massively multicore processors, such as Graphics Processing Units (GPUs), provide, at a comparable price, a one order of magnitude higher peak performance than traditional CPUs. This drop in the cost of computation, as any order-of-magnitude drop in the cost per unit of performance for a class of system components, triggers the opportunity to redesign systems and to explore new ways to engineer them to recalibrate the cost-to-performance relation.

This project explores the feasibility of harnessing GPUs' computational power to improve the performance, reliability, or security of distributed storage systems. In this context, we present the design of a storage system prototype that uses GPU offloading to accelerate a number of computationally intensive primitives based on hashing, and introduce techniques to efficiently leverage the processing power of GPUs.

We evaluate the performance of this prototype under two configurations: as a content addressable storage system that facilitates online similarity detection between successive versions of the same file and as a traditional system that uses hashing to preserve data integrity. Further, we evaluate the impact of offloading to the GPU on competing applications' performance. Our results show that this technique can bring tangible performance gains without negatively impacting the performance of concurrently running applications.

─────────── ◆ ───────────

## 1 INTRODUCTION

The development of massively multicore processors has led to a rapid increase in the amount of computational power available on a single die. There are two potential approaches to make the best use of this ample computational capacity available through massive parallelism: one is to design applications that are inherently parallel, while the other is to enhance the functionality of existing applications via ancillary tasks that improve an application's behavior along dimensions such as reliability and security.

While it is possible to expose the parallelism of existing applications, a significant investment is needed to refactor them. Moreover, not all applications offer sufficient scope for parallelism. Therefore, in the context of this project, we look into the second approach and investigate techniques to enhance applications along non-functional dimensions. Specifically, we start from the observation that a number of techniques that enhance the reliability, scalability and/or performance of distributed storage systems (e.g., erasure coding, content addressability [1, 2], online data similarity detection [3], integrity checks, digital signatures) generate computational overheads that often hinder their use on today's commodity hardware. We consider the use of Graphics Processing Units (GPUs) to accelerate these tasks, in effect using a heterogeneous massively multicore system that integrates different execution models (MIMD and SIMD) and memory management techniques (hardware and application-managed caches) as our experimental platform.

This project advocates the feasibility and evaluates the performance gains of building a GPU-accelerated storage system. We start by building a programming library, *HashGPU*, to accelerate hashing-based primitives which, although computationally demanding, are often used in storage systems (Section 2). We then quantify the end-to-end benefits of integrating GPU-offloading in a complete storage system (Section 3). To this end, we have prototyped a distributed storage system which integrates our *HashGPU* library with the *MosaStore* content-addressable storage system (Section 3). Most of the integration challenges are addressed by *CrystalGPU*, a generic runtime layer that optimizes task execution on the GPUs (Section 3.2.3). Our experimental evaluation (Section 4) demonstrates that the proposed architecture enables significant performance improvements compared to a traditional architecture that does not offload compute-intensive primitives.

The design approach our project has advocated, that is, exploiting GPUs as storage system accelerators, has recently been adopted by others in the storage systems community. For example the PTask project [4] proposes an OS abstraction for GPU management and demonstrates the PTask framework efficiency by building an encrypted storage system that offloads the encryption computation to GPUs. Further, Shredder [5] project adopts a similar design and optimization techniques proposed in this work to build a deduplicated storage system that supports backup systems and incremental computation.

*The contribution of this stream of work* is fourfold:
- First, we demonstrate the viability of employing massively multicore processors, GPUs in particular, to support storage system services. To this end, we evaluate, in the context of a content-addressable distributed storage system, the throughput gains enabled by offloading hash-based primitives to GPUs. We provide a set of data-points that inform the storage system designers' decision whether exploiting massively multi-core processors to accelerate the storage system operations is a viable approach for particular workloads and deployment environment characteristics.
- Second, we empirically demonstrate, under a wide set of workloads and configurations, that exploiting the GPU computational power enables close to optimal system performance; that is, employing additional compute power hashing computation will only result in minimal additional performance gains.
- Third, we shed light on the impact of GPU offloading on competing applications running on the same node as the

• *The authors are with the Electrical and Computer Engineering Department, The University of British Columbia, 2332 Main Mall, Vancouver, BC V6T 1Z4, Canada. E-mail: samera@ece.ubc.ca.*

distributed storage middleware. We focus on two issues in particular: First, while employing a GPU holds the potential to accelerate computationally intensive operations, the need to transfer the data back and forth to the device adds a significant load on a shared critical system resource, the I/O subsystem. Our experiments demonstrate that this added load does not introduce a new bottleneck in the system. Second, we quantitatively evaluate the performance impact on concurrently running compute- and IO-intensive applications.

- Finally, we introduce techniques to efficiently leverage the processing power of GPUs. GPUs' support for general-purpose programming (e.g., NVIDIA's CUDA [6]) reduces the effort needed to develop applications that benefit from the massive parallelism GPUs offer. However, significant challenges to efficiently use GPUs remain. To address these challenges, we have designed and integrated *CrystalGPU*, an independent runtime layer that efficiently manages the interaction between the hosted application and the GPU. *CrystalGPU* transparently enables the reuse of GPU buffers to avoid memory allocation overheads, overlaps the data transfer with computation on the GPU, and enables the transparent use of multiple GPU devices.

**The focus of this paper and relationship to authors' past work**. We have previously demonstrated that GPUs can accelerate the computation of hash-based primitives [7], and evaluated the end-to-end performance gains of one, comparably simpler to offload, hashing primitive: fixed size block hashing [8].

This paper, however, not only presents our experience over the last four years while working on this project, it further investigates the system-level integration challenges, and extends the exploration of the design space by extending our prototype storage system to use deduplication based on content-based chunking. While content-based chunking is significantly more computationally intensive than fixed block hashing primitive, investigated in our previous work [7], hence has a higher potential to benefit from GPU acceleration; it is stream orientated, a characteristic that complicates offloading it to the GPU which best operates on batched computations. Finally, this paper sheds light on a key opportunity issue often overlooked by other work that explores opportunities for GPU offloading: Given a fixed power or dollar budget, which configuration brings higher benefits, adding an extra CPU or a GPU to the system?

**Context**. Although this work focuses on hash-based primitives, we argue that the proposed approach can be extended to other computationally intensive routines that support today's distributed systems like erasure coding [9], compact dataset representation using Bloom filters [10] or Merkel trees, as well as data compression [11], deduplication [12], and encryption/decryption. To this end, parallel algorithms for these primitives exist, and can benefit from GPU support (e.g., parallel Reed-Solomon coding [13], parallel compression algorithm, and parallel security checking [14]).

Additionally, offloading can support an active-storage design for specialized storage systems that focus on, for example, content-based image retrieval [15]. In this context, data-parallel processing can be offloaded to the GPU to enable significant performance improvements, provided that the computation performed per byte of input data is sufficiently high to amortize the additional GPU overheads.

It is worth noting that multiple computational cores have been used to improve the performance of some security checks such as on-access virus scanning, sensitive data analysis, and taint propagation [14]. In fact, without parallel execution, the overhead of executing some of these tasks is prohibitive. The spirit of this work is similar: without support for parallel execution, many storage system optimizations cannot be deployed due to their associated overheads.

Recent commodity GPUs like the NVIDIA GeForce GTX 480 GPU (480 cores at 1.4 GHz) offer a ten-fold higher peak computational rate than Intel processors in the same price range. *This drop in the cost of computation, as any order-of-magnitude drop in the cost per unit of performance for a class of system components, triggers the opportunity to redesign systems and to explore new ways to engineer them to recalibrate the cost-to-performance relation.* Our evaluation of a GPU-accelerated storage system shows that GPUs can be a cost-effective enhancement to high-end server systems.

Further, using GPUs aligns with current technological trends: First, while multi-core CPUs currently do not offer this level of parallelism, they are expected to provide similar levels of parallelism over the next decade and support some of the techniques prototyped in a GPU context. Second, future massively multicore processors will likely extend the scope of heterogeneity already present in existing processors (e.g., AMD Fusion, Intel MIC, and IBM's Cell BE); they will likely integrate heterogeneous cores, multiple execution models (e.g., SIMD/MIMD core blocks), and non-uniform memory architectures [16, 17].

## 2 BACKGROUND

A number of primitives commonly used in distributed storage systems generate computational overheads that set them apart as potential bottlenecks in today's systems, which increasingly employ multi-Gbps links and/or are built on top of high throughput storage devices (e.g., SSDs [18]).

For instance, on-the-fly data compression offers the possibility to significantly reduce the storage footprint and network effort at the expense of performing additional computations. However, the compression throughput can be prohibitively low (4MBps for *bzip2* and 16MBps for *gzip* in our experiments on a 2.4Ghz Intel Core2 quad core processor) not only for high-performance computing scenarios but also for desktop deployments. Similarly, while encryption enables clients to securely store their data at an untrusted server, its throughput (between 38MBps and 157MBps in our experiments depending on the encryption algorithm used) can be lower than the client's network/disk throughput. Similarly, erasure codes' encoding throughput is limited by their computational complexity (for example, RAID6 systems using Reed Solomon codes achieve lower than 60MBps coding throughput using all cores of an Intel Core2 Quad processor [19]).

We aim to understand the viability of offloading these and other data-processing intensive operations to a GPU to dramatically reduce the load on the source CPU and enhance overall system performance. In this context, we focus on hashing-based primitives which are widely used in storage systems. This section highlights their computational overheads (2.1), and presents a brief overview of NVIDIA GPU's architecture (2.2).

## 2.1 Use of Hashing in Storage Systems

Hash-based primitives are commonly used in storage systems as building blocks to provide content addressability, data integrity checks, data similarity detection, compact set representation, and support for incremental computation (i.e. avoid processing regions of input that has already been processed). In this context, there are two main uses of hashing:

*Direct Hashing.* This technique computes the hash of an entire data block. In systems that support content addressability [1, 2, 12], data blocks are identified based on their content: data-block identifiers are simply the hash value of the data. This approach provides a number of attractive features. First, it provides a flat namespace for data-block identifiers and a naming system which simplifies the separation between file- and data-block metadata. Second, it enables probabilistic detection of similar data-blocks by comparing their hash value [1, 3]. Finally, it provides implicit integrity checks. The hash computation, however, imposes overheads that may limit performance, especially in two common scenarios. First, for workloads with frequent updates, changing a few bytes in the data may require rehashing multiple data blocks. Second, in systems that use large blocks (e.g., GoogleFS [20] uses 64MB blocks), computational overhead of hashing becomes a non-negligible component (possibly the largest) of a write/update operation.

*Content-based chunking.* The techniques mentioned above require dividing large files into multiple blocks. To this end, two approaches are possible: fixed-size blocks, and detecting block boundaries (i.e., the markers for blocks' start and end) based on content. This is done by hashing a large number of overlapping data windows inside the data-block and declaring a block boundary when the hash value matches a predefined value. Low-Bandwidth File System (LBFS) [3] and JumboStore [21] both adopt this latter technique, which we dub content-based chunking.

When used for similarity detection, these two techniques offer a tradeoff between computational overheads and the similarity ratio detected: unlike fixed-size blocks, detecting block boundaries based on content is stable to data deletions/insertions, yet significantly more computationally intensive, hence leading to lower throughput. In fact, the limited throughput of content-based chunking is the main reason its original proponents [3] recommend its use in storage systems supported by low-bandwidth networks. On an Intel Core2 Duo 6600 2.40GHz processor, and using data from a checkpointing application, our content based chunking implementation offered 7 to 51MBps throughput, and similarity detection rates between 82% and 37% respectively, depending on the configuration. Throughput in this range is clearly a bottleneck when using Gbps network links or fast disks.

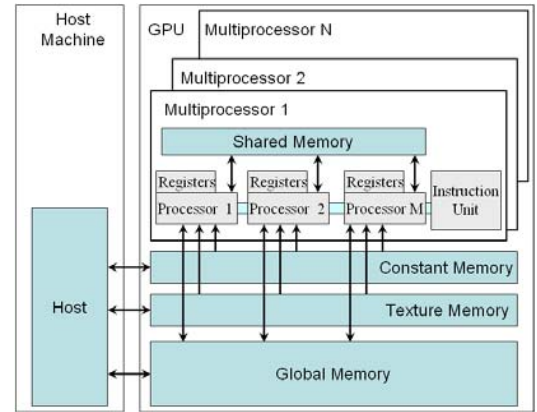

Figure 1. GPU Schematic Architecture.

## 2.2 GPU-Related Background

This section presents a brief overview of GPU architecture, programming model, and typical application structure. While we focus on NVIDIA's architecture and programming environment (the Compute Unified Device Architecture (CUDA) [6]) similar issues emerge with other GPU vendors or programming environments.

At a high-level view (Figure 1), NVIDIA's GPUs are composed of a number of SIMD multiprocessors. Each multiprocessor incorporates a small but fast *shared memory* (16 to 48KB). All processors in the multiprocessor have direct access to this memory. Additionally, all multiprocessors have access to three other device-level memory modules: the *global*, *texture*, and *constant* memory modules, which are also accessible from the host. The global memory supports read and write operations and it is the largest (with size ranging from 256MB to 4GB). The texture and constant memories are much smaller and offer only read access to GPU threads. Apart from size, the critical characteristic differentiating the various memory modules is their access latency. While accessing the shared memory takes up to four cycles, it takes 400 to 800 cycles to access global memory. Consequently, to achieve maximum performance, applications should maximize the use of shared memory and processor registers.

Table 1: The processing stages of a GPU task.

| Stage | Operations performed |
|---|---|
| (1) Pre-processing | Device initialization; memory allocation on the host and device; task-specific data preprocessing on the host. |
| (2) Data Transfer In | Data transfer from host's memory to device global memory. |
| (3) Processing | Loop through: (3.1) data transfer from global GPU memory to shared memory [optional]; (3.2) Task's 'kernel' processing; and (3.3) Result transfer back to global memory. |
| (4) Data Transfer Out | Output transfer from device to global memory to the host memory. |
| (5) Post-processing | Task-specific post processing on the host; resource release [optional]. |

A GPU task is composed of five stages (Table 1): preprocessing; host-to-device data transfer; kernel processing; device-to-host results transfer and post-processing. As Table 1 indicates, for a data-parallel application, the processing step is usually repeated until all input data is processed.

## 3 SYSTEM DESIGN

This section first discusses the design challenges of a GPU-accelerated storage system (§3.1), and then details our storage system design (§3.2).

### 3.1 Design and Integration Challenges

Efficient offloading to an accelerator (e.g., a GPU) generally implies executing the accelerated primitive in a different memory space. Support for efficient data transfers and thread management techniques are crucial to preserve the benefits of offloading. Additionally, accelerator-specific concerns emerge: for example, in case of a GPU, extracting the target primitive parallelism, and, equally important, employing efficient memory, and thread management techniques. Improper task decomposition, memory allocation and management, or data transfer scheduling can lead to dramatic performance degradation [22].

Apart from the above performance-related issues there are two additional areas to consider: minimizing the integration effort with an existing code (in our case a file system) and preserving the separation of concerns between application-related issues, accelerator logic, and resource allocation issues related to the accelerator.

Overall, seven areas of concern exist. We present them below starting with global, system-level concerns that impact all layers of our design, then continuing with issues that are specific to the GPU runtime management system, and finalizing with issues specific to the accelerated function

- *Minimizing the integration effort*. The volume of coding required to enable GPU-offloading should not dwarf the potential performance gains. Ideally, the primitives offloaded are stateless and offloading is simplified by preserving the API and the semantic of the original implementation on a CPU. Even in this case, however, there are two factors for additional complexity, as we shall argue: first, memory allocation which must be done at the device level, and, second, the asynchronous nature of task execution management on the GPU as well as other device-management operations. Our design hides these complexities under simple, intuitive interfaces. (§3.2.1 our solution based on layering).

- *Separation of concerns*. The integrated design should preserve the separation between the logic of the accelerated primitive(s) and the runtime management layer required for efficient resource management at the accelerating device. The main driver for this requirement is facilitating the addition of new primitives that will benefit from offloading through the same runtime layer. Our layered design (Figure 2, §3.2.1) and the generic task-management functionality offered by the *CrystalGPU* layer (§3.2.3), address this concern.

- *Batch oriented computation*. A system adopting GPU offloading will deliver higher performance gains if it can support batch-oriented computation for the offloaded functions; as GPUs best support batch computations. This generally involved module decoupling and making the calls to the accelerated primitives asynchronous. This change represents a significant challenge as it is often hard to obtain without further complicating the first two concerns: complicating integration or breaking the system's design.

- *Hiding data transfer overheads*. For streaming applications, that is, for applications that send multiple GPU-tasks back to back, overlapping the transfer of input data to the GPU with the computation step of a previous task can hide the data transfer overhead. We explore these optimizations through the *CrystalGPU* layer (§3.2.3).

- *One-copy host to device data transfers*. Data transfers to and from the device require DMA transfers from (respectively to) non-pageable host memory. If an application presents data residing in pageable memory for transfer to the GPU then the CUDA runtime library first allocates a new buffer in non-pageable memory, copies the data to this new buffer, and finally launches the DMA transfer. Thus, to avoid these additional overheads (data-copy and new buffer allocation), the application should present data in buffers allocated in non-pageable memory.

- *Hiding memory allocation overheads*. Additionally, since allocating memory in non-pageable memory is an expensive operation, the application should reuse buffers to the extent possible. To this end we have added a memory management module to the CrystalGPU layer (§3.2.3). This module offers a CrystalGPU-specific implementation of `malloc` and `free` function calls and manages non-pageable memory buffers used through the application run (these buffers are allocated at the application initialization).

- *Simplifying the use of multiple GPUs in multi-GPU systems*. CUDA provides only a low-level API for using multiple GPUs by the same application. Managing and balancing the load between multiple GPUs significantly increases application complexity. To circumvent this limitation, and benefit from multi-GPU systems, *CrystalGPU* provides an API that enables transparent and balanced use of multiple GPUs. (§3.2.3).

- *Efficient use of shared memory*, the fastest, yet the smallest, GPU memory module. Three reasons make the efficient use of shared memory challenging. First, shared memory is often small compared to the volume of data being processed. Second, shared memory is divided into banks and all read and write operations that access the same bank are serialized (i.e., generate bank conflicts), hence, reducing concurrency. Consequently, to maximize performance, the concurrent threads should access data on different banks. The fact that a single bank does not represent a contiguous memory space increases the complexity of efficient memory utilization. Finally, while increasing the number of threads per multiprocessor helps hiding global memory access latency, this does not directly lead to a linear performance gain. The reason is that increasing the number of threads decreases the amount of shared memory available per thread.

To efficiently use the shared memory without increasing programming complexity, we have implemented a shared memory management mechanism (§3.2.2).

### 3.2 A GPU-accelerated Storage System Prototype

The prototyped distributed storage system integrates three main components: the *MosaStore* storage system, *HashGPU* library, and *CrystalGPU* the accelerator runtime management system. The integrated system is presented in Figure 2 and Figure 3. We integrated *MosaStore* with *HashGPU* and *CrystalGPU* such that the compute-intensive hash-based processing required to support content addressability is outsourced to a GPU. The rest of this section details the design of each of these three components (§3.2.1 to §3.2.2) and the role of each of them in the final system prototype (§3.2.4).

*3.2.1 MosaStore*
*MosaStore* is a highly configurable storage system prototype that can be configured to optimize its operations to exploit specific workload characteristics [23]. More relevant to this study, the current version of *MosaStore* can be configured to work as a content addressable storage system.

*MosaStore* employs an object-based distributed storage system architecture (similar to GoogleFS). Its three main components are (Figure 2): a centralized metadata manager, the storage nodes, and the client's system access interface (SAI), which uses the FUSE [24] kernel module to provide a POSIX file system interface.

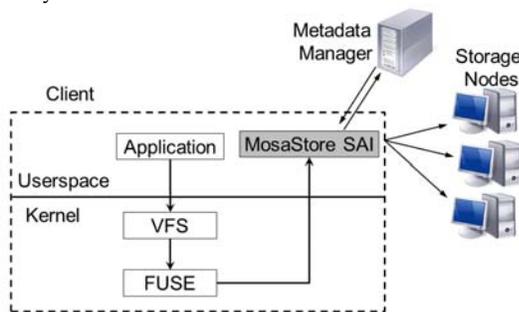

Figure 2. System architecture. At the client node, we use the FUSE Linux module to implement a user-level filesystem (MosaStore SAI). The Linux virtual file system (VFS) calls, through FUSE, user-defined callbacks, that implement MosaStore file system functionality. Note that the GPU is needed only on the client machines. Figure 3 presents in detail the structure of the MosaStore SAI that integrates the GPU.

Each file is divided into blocks that are stored on the storage nodes. The metadata manager maintains a block-map for each file which contains the file's blocks information including the hash value of every block. The SAI implements the client-side content-based addressability mechanisms (which can be configured to use either fixed-size or content-based chunking). This generates the hashing workload that traditionally is executed on a CPU or can be offloaded to the GPU via *HashGPU* module.

To write to *MosaStore*, the SAI first retrieves the file's previous-version block-map from the manager, divides the new version of the file into blocks, computes the hash value of each block, and searches the file's previous-version block-map for equivalent block hash values. The SAI stores only the blocks with no match at the storage nodes, saving considerable storage space and network effort. Once the write operation is completed (as indicated by the `release` POSIX call) the SAI commits the file's block-map including the blocks' hash values to the metadata manager. The architecture of the MosaStore SAI and its use of GPU offloading are presented in detail in Figure 3 and its legend.

*3.2.2 HashGPU*
*HashGPU* implements, on a GPU, the two hashing primitives often used in storage systems (presented in §2.1): *direct hashing*, i.e. hashing of a large block of data, with size ranging from kilobytes to megabytes or more, and *content-based chunking*, i.e., repeatedly hashing a small sliding window through a data stream to detect chunk/block boundaries. The rest of this section presets the high-level design of these two modules, and GPU specific performance considerations related to porting code to a GPU.

*Direct Hashing.* The most widely used hash functions follow the *sequential* Merkle-Damgård construction approach [25, 26]. In this approach, a large data block is split in fixed-size segments. The first segment is processed to produce a fixed-size output which is used as the input to the hashing function together with the next segment. The process continues sequentially until all segments are processed. This sequential construction does not allow multiple threads to operate concurrently to hash the data. To exploit GPUs' parallelism, *HashGPU* employs the *parallel* Merkle-Damgård construction: the sequential hash function is used as a building block to process multiple segments of data in parallel on the GPU. We note that the resulting hash is as strong as the original, sequentially built, hash, as demonstrated by Damgård [26].

*Content-based chunking.* To parallelize the computation of a large number of small hashes drawn from a large data block, we hash in parallel all the small blocks and aggregate the result in a buffer. The challenge in implementing this module lies in the memory management to extract maximum performance. Note that the input data is not divided into smaller blocks as the previous case. The reason is that the input data for each thread may overlap with the neighboring threads. As there are no dependencies between the intermediate hashing computations, each computation can execute in a separate GPU thread. For an in-depth description we refer the reader to our previous publication [7].

The design of the two modules is similar: Computing the hash (direct hashing of content-based chunks) is carried in five stages (as described in §2.1). Once the input data is transferred from the CPU, it is divided into segments that are hashed in parallel. Each segment is hashed using a standard sequential hashing function (MD5 in all our experiments), note that the segment may overlap in case of sliding window hashing. The result is placed in a single output buffer. The output buffer is further processed to produce the final result: direct hashing hashes the buffer to produce the result hashing value, and sliding window hashing checks the buffered hash values to detect block boundaries. Two aspects are worth mentioning. First, as there are no dependencies between the intermediate hashing computations, each computation can execute in a separate GPU thread. Second, this design uses the CPU to compute the last step: in direct hashing, the CPU is used to compute final hash of the intermediary hashes; while in content-based chunking, the CPU is used to check the hash values and decide on block boundaries. The reason for using the CPU in the last stage is that efficiently synchronizing all running GPU threads is not possible [6].

*Optimized Memory Management.* Although the design of the hashing module is relatively simple, optimizing for performance is a challenging task. For example, one aspect that induces additional complexity is maximizing the number of threads to extract the highest parallelism (around 100K threads are created for large blocks) while avoiding bank conflicts and maximizing the use of each processors' registers. To address this issue, we have implemented a shared-memory management subsystem that:
- *Reduces memory access latency* by allocating to each thread a

fixed-size workspace located in shared memory. Additionally, to avoid bank conflicts, the workspaces of threads that are co-scheduled are allocated on separate shared memory banks. When a thread starts, it copies its data from the global memory to its shared memory workspace, hence avoiding subsequent accesses to the slower global memory.

- *Reduces the complexity of the programming task.* As the workspace allocation technique just described makes programming more complex; since a shared memory bank is not mapped to a contiguous memory address space. To reduce programing complexity the memory management subsystem abstracts the shared memory to allow the thread to access its workspace as a contiguous address space.

In effect, the shared memory management subsystem reduces memory access latency by avoiding bank conflicts while reducing the programming effort by providing a contiguous memory address abstraction. Further, the shared memory management subsystem can bring benefits even to the latest GPUs with hardware–controlled caches as it performs intelligent data placement to avoid bank conflicts.

*3.2.3 CrystalGPU*

Although *HashGPU* optimizes the processing of a single data block, there is still room for improvement. Let us consider a workload that is a stream of hashing computations. In this case, three opportunities for performance improvement exist: input/output buffer reuse to amortize buffer allocation overhead, ability to harness multiple GPUs on multi-GPU systems, and exploiting the CUDA2 [6] capabilities to overlap the data transfer of one block with the computation of a previous block.

To exploit these opportunities, we developed *CrystalGPU*, a standalone abstraction layer that exploits these three opportunities. *CrystalGPU* runs entirely on the host machine as a management layer between the application and the GPU native runtime system. *CrystalGPU* manages the execution of GPU operations (e.g., transferring data from/to the GPU and starting computations on the GPU) and includes a memory management layer that enables the reuse of non-pageable memory.

At a high level, the metaphor *CrystalGPU* offers is that of a task-management environment for GPU tasks. Our design aims for (i) *generality*, to facilitate the support of a wide range of GPGPU applications, (ii) *flexibility*, to enable deployment on various GPU models while hiding the heterogeneity in task management across models, and (iii) *efficiency*, to maximize the utilization of GPU resources (processing units and I/O channels). We achieve these goals via an interface that is agnostic to the upper level application, and an internal design that avoids extra data copies and shared data-structure bottlenecks. Due to space limitations we do not present the detailed *CrystalGPU* design and API.

*CrystalGPU* design comprises a single driving module named the master. The master module employs a number of host-side manager threads, each assigned to manage one GPU device. The rationale behind this assignment is twofold. First, each manager thread is responsible for querying its device for job completion status, and asynchronously notifying the application, using the callback function, once the job is done. This allows the client to make progress on the CPU in parallel; further, it relieves the application from job execution state bookkeeping. Second, having a dedicated control thread for each GPU facilitates transparent multi-GPU systems support. As detailed below, the application submits a job to a shared outstanding queue, later the job is transparently handled by one of the manager threads in such a way that balances the load across GPUs. We note that this multithreaded design requires a multi-core CPU to enable maximum parallelism across the manager threads as well as the application's host-side threads. The application requests services from the framework by submitting jobs and waiting for callbacks. The status of a job is maintained by the master using three queues. First, the idle queue maintains empty job instances (with pre-allocated buffers). Second, the outstanding queue contains ready-to-run jobs submitted by the application, but not dispatched to the GPUs yet. Third, the running queue contains the jobs that are currently being executed on the GPUs. A manager thread is chosen in a round robin scheme to execute the next job in the outstanding queue.

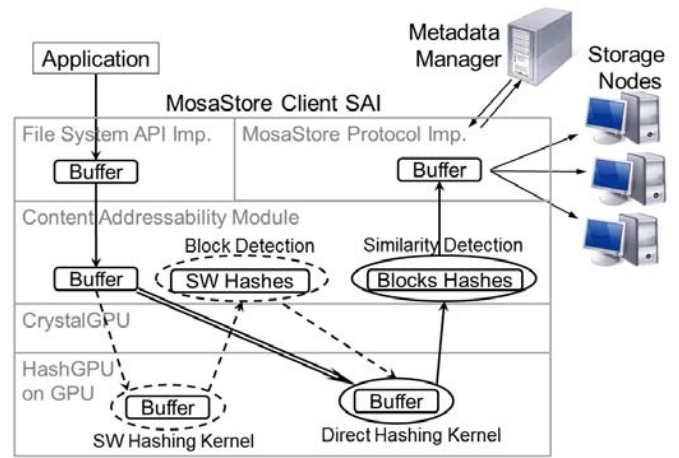

Figure 3. MosaStore SAI architecture and operations flow. The data written by the application is stored in a buffer. When the buffer is full, the content addressability module submits to *CrystalGPU* a request to execute the *HashGPU sliding window* (SW) hashing kernel on this data. The resulting sliding window hashes are used to detect the blocks boundaries. Upon detecting these boundaries, the content addressability module submits a second computation to *CrystalGPU* that uses, this time, the *HashGPU direct hashing* kernel to compute block hashes. Once these are received the content addressability module uses the blocks' hashes to detect similarity and decide if the any of the blocks are new and need to be transferred to the storage nodes. All buffer management and *HashGPU* kernel calls are managed by *CrystalGPU*. The dashed ovals and arrows represent operations that are only executed in content based chunking content addressability, and the double lined arrow represents the flow of operation in fixed size blocks content addressability.

*3.2.4 Integration and System Prototype*

Our prototype (Figure 2 and Figure 3) integrates the three aforementioned components into a storage system able to exploit the GPU computational power to support content addressability. The integration entailed changes to the storage system design to efficiently exploit the GPU.

Two GPU characteristics make harnessing their computational power challenging: the high data transfer overheads to/from the GPU, and the highly parallel GPU architecture. To match with these characteristics, workloads that best fit GPUs are ideally highly parallel (in the SIMD model) and involve a high computation-per-transferred-byte rate. Data-intensive stream-oriented workloads such as those generated by storage systems using content-based chunking are not an ideal match. Batching the streamed computations, however, can be used to mold these workloads to better exploit the GPU. The rest of this section discusses the main design and implementation changes for integrating the two GPU modules: fixed size blocks and content-based chunking.

*General Changes.* In addition to the module-specific changes detailed below we modified our original implementation of *HashGPU* library [27] to use the GPU as virtualized by *CrystalGPU*. To this end, we applied two changes: allocated the data buffers through *CrystalGPU*, and retrofitted a hash computation as a *CrystalGPU* 'task' (A task is *CrystalGPU's* abstraction for a unit of GPU computation and the associated data transfers). The *HashGPU* resulting library is able to harness optimizations available for a single hash operation as well as across a stream of hash operations.

*Fixed-Size Blocks Hashing Integration.* As fixed-size block hashing operate at a relatively larger granularity (and internally *HashGPU* treats one block as one batch), we have decided to ignore buffering optimizations, which made changing *MosaStore* to efficiently exploit the GPU for fixed blocks a relatively easy integration task. We simply modified *MosaStore's* SAI to use the *HashGPU* library to offload the hash computations to the GPU. This modification required two straightforward changes: First, all hash function calls needed to be changed to the new *HashGPU*-provided hash function (Since APIs are similar this effort is minor). Second, to facilitate optimized data transfers and buffer reuse during GPU calls, memory buffers needed to be allocated using the *HashGPU* library call instead of the standard memory allocation functions (e.g, `malloc`). The resulting storage system is able to offload the hash computation to the GPU.

These changes required changing only 22 lines of the original *MosaStore* implementation (over 18K lines of code). We estimate that integrating GPU offloading with other storage systems that make use of fixed size blocks hashing is equally straightforward.

*Content-based Chunking Integration.* Storage systems employ content-based chunking to detect block boundaries in newly written data; hence often sliding window hashes are computed on the data as it is written (often in chunk of 4KB in size). This design approach offers little opportunity for exploiting GPU power since the communication overhead is high and there is little parallelism to exploit. To better support GPU offloading we applied major changes to the *MosaStore* SAI to use content-based chunking in a batch oriented fashion. As data is written through the SAI to the file system the data is buffered. Once the buffer is full, it is submitted to *HashGPU* which computes the block boundaries of the buffered data. The newly discovered blocks are further processed by *HashGPU* direct hashing module for computing each block's hash value. When block boundaries do not align with buffer sizes care must be taken to transfer the leftovers to first block that is detected in the next buffer.

## 4 EXPERIMENTAL EVALUATION

This section evaluates the performance benefits of using GPUs as storage system accelerators. We evaluate component and integrated system performance using synthetic as well as real application workloads.

This section starts by highlighting the benefits brought by *CrystalGPU* when used in conjunction with *HashGPU* (§4.1). While we focus on this comparison, we also include essential information to highlight the performance offered by *HashGPU* when used independently. We refer the reader to our previous publication for a complete evaluation of *HashGPU* library [7]. Next, we investigate the following question: Given a fixed budget, which configuration can bring higher performance: adding an extra CPU or GPU to the system? (§4.2).

Finally, the evaluation focuses on the integrated system, and analyzes the performance improvements GPU support brings to the whole storage system (§4.3); demonstrates that with the configuration we propose we reap almost all possible potential gains additional computation can provide (§4.4); and studies the impact of offloading to the GPU on concurrently running applications (§4.5).

### 4.1 Evaluating the Performance Gains Enabled by CrystalGPU

To explain the performance gains enabled by the *CrystalGPU* layer, this section starts by presenting the overheads of the various GPU processing steps within the *HashGPU* library when used independently. We then evaluate the performance gains enabled by each of the optimizations enabled by *CrystalGPU*: buffer reuse, overlap of data transfers and computations, and transparent utilization of multiple GPUs.

The evaluation uses a machine with an Intel Xeon Quad-core 2.33 GHz, 16GB of memory, a PCI Express 2.0 x16 bus, and an NVIDIA GeForce GTX 480 GPU with 480 1.4 GHz cores and 1.5 GB of memory.

**The Overheads**. Figure 4 presents the proportion of the total execution time that corresponds to each of the stages outlined in Section 2.2 for the sliding window hashing module. These results show the major impact of memory allocation and copy-in stages on the total execution time: up to, 80-96% of the total execution time in sliding hashing module (Figure 4), and up to 70-90% in direct hashing module (presented in [7]).

Two opportunities exist to reduce the overhead of these two stages: first, the overhead of memory allocation stage can be avoided by reusing memory buffers. Second, copy-in overheads can be hidden, by overlapping data transfers and computation. Given the generic nature of these optimizations, *CrystalGPU* offers them in an application agnostic layer. The rest of this section presents an evaluation of the benefits of integrating *HashGPU* on top of *CrystalGPU*.

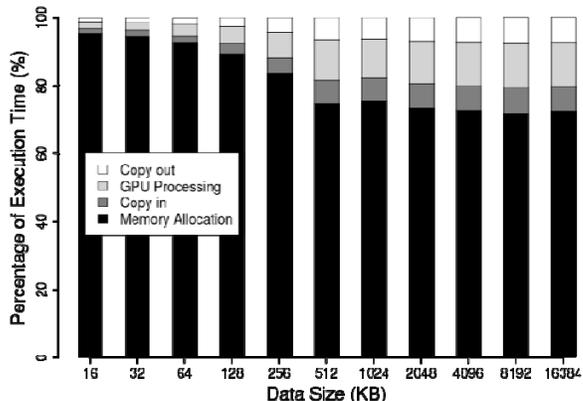

Figure 4: Percentage of total HashGPU sliding window hashing execution time spent on each stage without any optimization.

**The Performance Gains.** Figure 5 and Figure 6 show the average speedup obtained when using *HashGPU* on top of *CrystalGPU* for the sliding window hashing and direct hashing modules, respectively, to hash a stream of 10 data blocks. The baseline for speedup calculation is single core performance. In addition to speedup the figures also present the processing throughput for critical lines. The dual GPU configuration uses a second GPU card in the system: NVIDIA Tesla C2050 GPU with 448 1.1GHz cores and 3GB memory. While we use batches of 10 blocks our evaluation, not presented here for lack of space, shows that a batch of at least 3 blocks is needed to obtain close to maximal performance gains. We note that, to avoid cluttering the figures, we do not plot the standard deviation as it is significantly small for all experiments.

The experiment shows the effect of block size on the achieved speedup. For smaller block sizes, and irrespective of which optimization is enabled, the memory allocation and data transfer overheads overshadow the performance improvement in the computation (i.e., calculating the hash); conversely, for larger block sizes, the aforementioned overheads are amortized over the longer computation time, hence achieving better speedups compared to the CPU.

This experiment demonstrates that using all optimizations enabled by *CrystalGPU* leads to huge performance gains; up to 190x performance gains in sliding window hashing and 45x performance gains in direct hashing (for large data blocks), compared to up to 27x and 7x performance gains, respectively, provided by *HashGPU* alone. Further, while the original sliding window hashing performance lags behind the CPU performance for data blocks smaller than 64KB, and while the original *HashGPU* direct hashing performance lags behind the two socket CPU performance for all data sizes, adding *CrystalGPU* enables much higher performance gains.

The figures demonstrate that the buffer reuse optimization is able to amortize memory management costs; enabling, as a result, up to 100x sliding window hashing speedup for large data blocks. The overlap feature enables an additional speedup increase that corresponds to the portion of time spent on data transfer operations (to up to 125x). Using both GPUs available on the machine increases the speedup further (to up to 190x). Direct hashing results (Figure 6) show a similar pattern. Two contention points prevent dual GPU cards to achieve linear speedup in our case: First, the limited percentage of time spent in computation (compared to the overheads of memory allocation and I/O). Second, in the direct hashing case, post-processing stage is performed sequentially on the CPU.

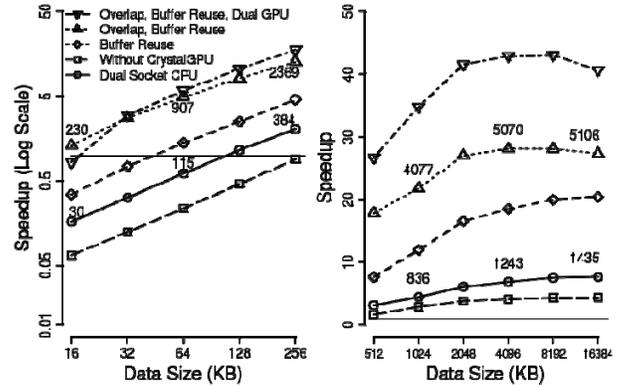

Figure 5. Achieved speedup for *sliding window hashing* for a stream of 10 jobs. Small data sizes in the left figure (logarithmic scale on Y axis), and large data size in the right figure (linear scale on Y axis). Values below 1 indicate slowdown. The baseline is the performance on a single core on Intel Xeon Quad-core 2.33 GHz. The numbers over the "*Dual socket CPU*" and "*Overlap, Buffer reuse*" datapoints indicate the achieved processing throughput in MBps.

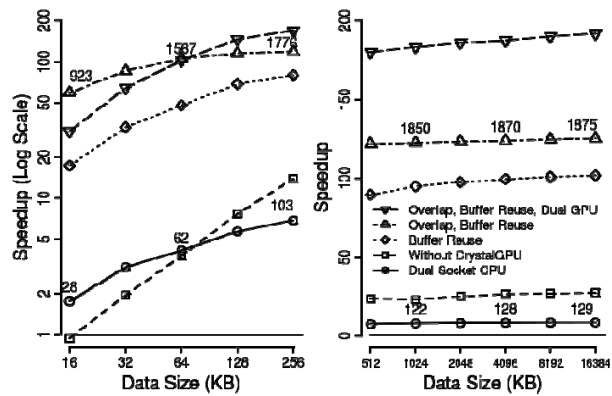

Figure 6. Achieved speedup for *direct-hashing* for a stream of 10 jobs. Small data sizes in the left figure (logarithmic scale on Y axis), and large data size in the right figure (linear scale on Y axis). Values below 1 indicate slowdown. The baseline is the performance on a single core on Intel Xeon Quad-core 2.33 GHz. The numbers over the "*Dual socket CPU*" and "*Overlap Buffer reuse*" datapoints indicate the achieved data processing throughput in MBps.

### 4.2 Add a CPU or a GPU?

To inform the system builders' decision, this section focuses on comparing the performance gain when extending the system with a second CPU versus a GPU card. We compare the performance of a system originally configured with an Intel Xeon Quad-core 2.33 GHz CPU and 16GB of memory when extended either by a second similar processor, or with an NVIDIA GeForce GTX 480 GPU with 480 1.4 GHz cores and 1.5 GB of memory.

Our CPU evaluation uses a multithreaded implementation of the content-based chunking module. Our evaluation (not presented here due to space limitations) shows that using 16 threads leads to the highest performance on the dual CPU system. Our GPU evaluation uses *HashGPU* on top of *CrystalGPU* as presented in the previous section.

Figure 5 and Figure 6 shows the speedup obtained when using an additional CPU (the solid line labeled "*Dual Socket CPU*") compared to using a single GPU card (the dotted line labeled "*Overlap, Buffer Reuse*"). The figure shows that the single GPU configuration achieves up to 125x sliding window hashing performance gains (relative to our baseline) compared to only 8x performance gain by the dual CPU configuration (a 15x relative performance gain). Further, the single GPU configuration achieves up to 28x direct hashing performance gains compared to only 8x performance gain by the dual CPU configuration (a 3.5x relative performance gain).

Further, we note that despite the 8x performance improvement of the dual CPU configuration, this configuration only achieves 129MBps sliding window hashing throughput, a throughput close to the 1Gbps network throughput and significantly lower than the throughput of high performance networks (often 10Gbps). Consequently, regardless of the level of similarity detected sliding window hashing on two CPUs will not bring performance benefits for 1Gbps system and will introduce a performance bottleneck for 10Gbps systems (the next section presents an end-to-end evaluation).

These results highlight that adding a GPU is a better fit for the hashing-based workloads due to its highly parallel architecture and high throughput memory system.

### 4.3 Integrated System Evaluation

This section evaluates the gains enabled by integrating the *HashGPU/CrystalGPU* modules with *MosaStore*. We evaluated the integrated system on a 22 nodes cluster of 2.33GHz Intel Xeon Quad-core CPU, 4GB memory nodes connected at 1Gbps. The storage system client machine has two CPUs of Intel Xeon Quad-core 2.33 GHz (8 cores in total), 16GB of memory, a PCI Express 2.0 x16 bus, and NVIDIA GeForce GTX 480 GPU. We present here averages over 10 runs. To avoid cluttering the figures, we do not plot the standard deviation as it is significantly small for all experiments.

We evaluate the system using two configurations: *Fixed size blocks* that uses direct hashing for similarity detection. The block size is 1MB, the default block size in MosaStore, and *Content based chunking* configuration, which uses sliding window hashing to detect block boundaries. The block size was 1.2MB on average (with minimum block size of 256KB and maximum block size of 4MB). The client SAI is configured to stripe the write operations to four storage nodes in parallel.

To explore the gains enabled by GPU offloading we compare the performance of the following three configurations of the storage system:
- *non-CA*, in which the content addressability module is disabled and all data is written directly to the storage nodes (i.e., without any hashing and similarity detection overheads).
- *CA-CPU*, in which all processing related to content addressability is done by the CPU (i.e., hashing of data blocks and hash comparisons). Compared to the previous configuration, this configuration exposes the performance impact of hashing data to support content addressability.
- *CA-GPU*, which uses the integrated *HashGPU/CrystalGPU* stack to offload the computation of hashes to the GPU. Compared to the previous configuration, this configuration exposes the gains brought by offloading computationally intensive operations to a GPU.

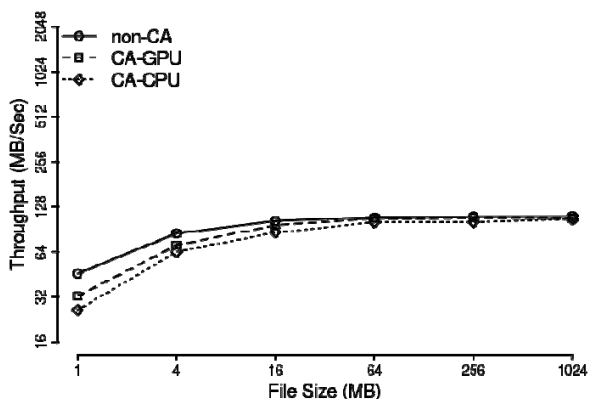

Figure 7: Average throughput while writing 40 different files back-to-back in the *fixed block* configuration. Note the logarithmic scale on the y-axis.

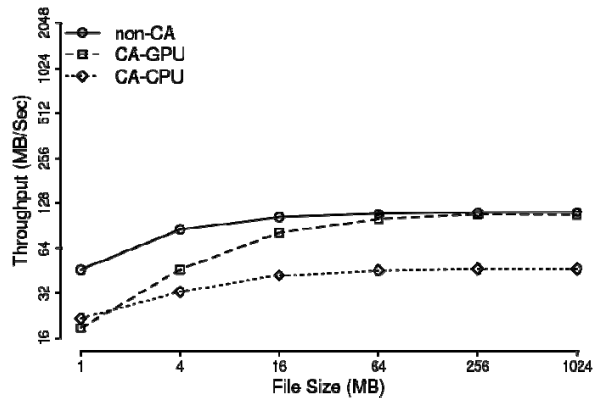

Figure 8: Average throughput while writing 40 different files back-to-back in the *content based chunking* configuration. Note the logarithmic scale on the y-axis.

Note that the performance of the system when using content addressability varies depending on the degree of data similarity present in the workload. To evaluate the entire performance spectrum we use the following three workloads:
- *Different*: The first workload consists of writing a set of completely different files. This workload exposes all overheads as all data need to be hashed and transferred across the network to the storage nodes. Moreover, no similarity can be detected between writes, which implies no opportunity to reduce space or bandwidth usage. Exploring this workload has a second advantage: the performance comparison holds for systems that use hashing for other goals than content addressability (e.g., for data integrity checks only).
- *Similar*: The second workload represents the other end of the spectrum: it exposes an upper bound for the performance gains that can be obtained using content addressability and maximizes the hash-computation overheads in relation with other storage overheads. When the files are identical, data is transferred only once across the network yet similarity detection overheads still exist.
- *Checkpoint*: Finally the third workload represents a real application data. This third workload involves 100 successive data checkpoint images, taken at 5 minute intervals for the BLAST [28] application using BLCR [29] checkpointing library (the average checkpoint size is 264.7MB). Fixed size blocks similarity detection detects 21-23% similarity on average between successive checkpoint images, depending on the block size, while content based chunking detects 76-90% similarity depending on the block size.

*Results for the "Different" Workload.* Figure 7 presents the write throughput with the *different* workload with fixed size blocks configuration. The figure shows that under workloads with completely non-similar data, the *non-CA* configuration achieves the highest throughput while the *CA-CPU* and *CA-GPU* configurations lag behind for small files. The reason is that with completely non-similar files, the similarity detection mechanism increases the write operation's overhead without bringing any performance gains (expected by reducing the amount of data that need to be transferred).

Figure 8 presents the write throughput under the content based chunking configuration. The figure shows that the *CA-CPU* and *CA-GPU* configuration lag behind *non-CA* in almost all file sizes. We note that content based chunking configuration using dual CPUs is caped at 46Mbps (significantly lower than the network throughput of a 1Gbps NIC card) highlighting that content based chunking introduces a new performance bottleneck for the write operation (this is also the case with the *similar* workload presented next).

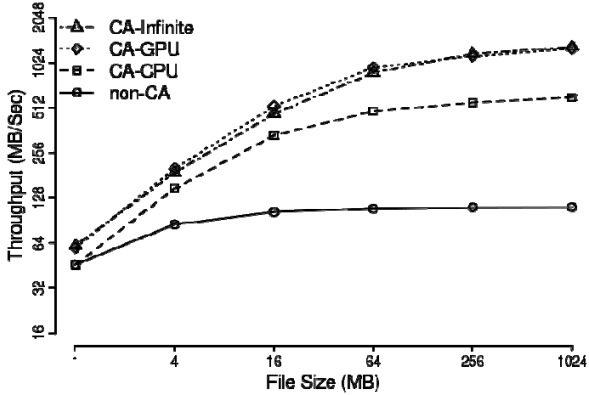

Figure 9. Average throughput while writing the same file 40 times back-to-back in the *fixed block* configuration. We discuss the CA-Infinite configuration in the next section. Note the logarithmic scale on the y-axis.

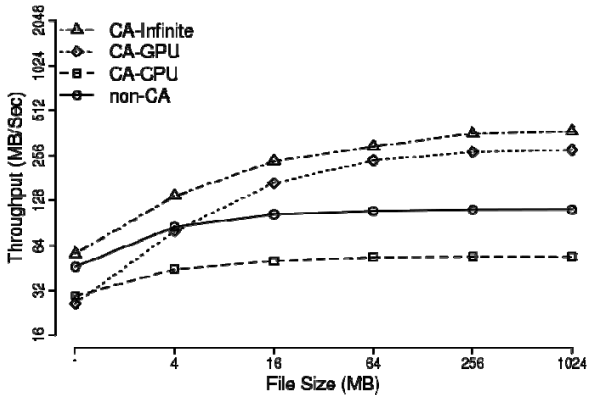

Figure 10: Average throughput while writing the same file 40 times back-to-back in the *content based chunking* configuration. We discuss the CA-Infinite configuration in the next section. Note the logarithmic scale on the y-axis.

*Results for the "Similar" Workload.* Figure 9 presents the write throughput with the *similar* workload with fixed size blocks configuration. The figure shows that under workloads with completely similar data, *MosaStore* with *HashGPU* significantly outperforms the CPU version and achieves over two times higher throughput for files larger than 64MB. This is because the similar workload is compute bound: only the first file needs to be transferred over the network to the storage nodes, while the computationally intensive hash computations will detect that the other files are similar. Under this workload *HashGPU* enables doubling the system throughput by accelerating the hash computations.

Figure 10 presents the write throughput for the *similar* workload with content based chunking configuration. We note that the CPU version achieves lower throughput than *non-CA* configuration. This is because the content based chunking approach adds significant computation overhead that introduces a new bottleneck in the system. On the other hand, MosaStore with *HashGPU* significantly outperforms the CPU and *non-CA* versions and achieves over 4.4x and 2.1x, respectively, higher throughput for files larger than 64MB.

This result validates the hypothesis that while content based chunking can reduce the transferred data size, its computation overhead on the CPU hinders its use in high performance computing system. Offloading this computational step to GPU not only eliminates this bottleneck but brings significant performance gains.

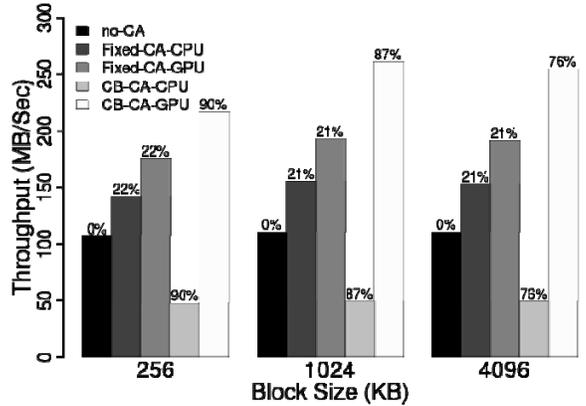

Figure 11. Average throughput while writing 100 BLAST checkpoints back-to-back while varying the block size. "*Fixed*" denoted evaluation with the fixed block size configuration while "*CB*" denotes the content based chunking configuration. The content based chunking configuration was tuned to produce average chunks sizes close to the sizes indicated on x-axis. The numbers on top of the bars denote the average similarity percentage detected using the configuration.

*Results for the "Checkpointing" Workload.* Figure 11 presents the write throughput with the checkpoint workload for all system configurations *while varying the block size*. These results lead to the following observations:

- First, offloading to GPU always offer sizeable benefits compared to a system with dual CPU configuration. offloading to the GPU enables about 1.3x write throughput improvement with fixed block sizes and up to 5x write throughput improvement with content based chunking compared to a content addressable system that does not offload (i.e. uses the CPU).
- Second, content based chunking configuration achieves the highest throughput when offloaded to GPU; up to 5x higher than CPU configuration and 2.3x than a system without content addressability. Although content based chunking adds higher overhead in detecting blocks boundaries, it detects 3-4x more similarity across checkpoint images, significantly reducing the amount of data that needs to be transferred, hence offering higher throughput in the GPU offloading case.
- Third, content based chunking using dual CPUs achieves the lowest throughput (49MBps) despite the high similarity ratio detected. This is because the system is bottlenecked by the content based chunking mechanism on the CPUs.
- Fourth, the speedup offered by offloading to the GPU increases as the block size increases (as evaluated in sections 4.1 and 4.2). This is the reason the system that offloads to the GPU performs increasingly better for 1MB, and 4MB blocks.
- Finally, the block size offers a performance tuning knob: small block sizes enable detecting higher similarity, but increase the system overhead (for transferring larger number of data blocks, and similarity detection). This tradeoff is clearer with content based chunking configuration when offloaded to the GPU. The configuration achieves the highest throughput with block sizes close to

1MB although higher similarity is detected using 256KB blocks, and although larger blocks (i.e 4MB) are expected to reduce data transfer overheads.

In summary, for workloads that have data similarity, exploiting GPU's computational power can bring tangible performance benefits to content addressable storage systems. Further, in systems that use hashing only for data integrity checks (as indirectly evaluated by the *different* workload), hashing can be efficiently offloaded to the GPU.

### 4.4 What would the system performance be if we had infinite compute power?

While our results (Figure 9 and Figure 10) show that significant performance gains can be achieved by offloading the hash computation to the GPU, it is important to estimate the practicality of further investing in CPU/GPU hardware or software optimizations to obtain higher throughput. To answer this question, we experimented with a hypothetical storage system configuration (CA-Infinite in Figure 9 and Figure 10) that represents a system with infinite compute power for computing hash functions (sliding window hashes or direct hashes). This configuration uses an oracle that 'computes' the hash function instantly (thus emulating infinite compute power to compute hashes). The results in Figure 9, (which presents the throughput of the system when using the similar workload with fixed block configuration) show that CA-GPU performance is almost equivalent to optimal (as represented by CA-Infinite plot line). Further, the results in Figure 10, (which presents the throughput of the system using content based chunking) show that CA-GPU performance is close to optimal: the throughput loss is lower than 50% for files smaller than 16MB, and lower than 25% for larger files. Thus, we can argue that offloading to the GPU, which in this configuration and workload combination enables up to 2.3x higher system throughput, will offer close to the maximal benefits that could be obtained by accelerating the hash computation on standard commodity clusters.

### 4.5 The Impact on Competing Applications

While the previous section has demonstrated that offloading computationally intensive primitives to the GPU can improve the system's throughput, the impact of this approach on the overall client system performance still needs to be evaluated. On the one hand, offloading frees CPU cycles on the client system. On the other hand, offloading might add a significant load on the client's kernel and the I/O subsystems to handle data transfers to and from the device. Consequently, this technique may negatively impact applications that are running concurrently, especially those with high I/O load.

This section evaluates the impact of the proposed approach on concurrently running applications. We use two workloads to reproduce the diversity of applications' possible resource usage pattern: a compute bound application (we use a multi-threaded prime number search application), and an I/O-bound application (we use the compilation of the Apache web server v2.2.11 as a representative I/O-bound application, which particularly stresses the disk I/O channel). Further we evaluate the system using *MosaStore* with fixed size block configuration, as content based chunking approach using CPUs clearly adds significant computational overhead to the system and is a less viable option in performance centric HPC systems (Section 4.3).

The evaluation uses a client machine with an Intel quad-core 2.66 GHz processor, PCI Express 2.0 x16 bus, and NVIDIA GeForce 9800GX2 GPU.

To measure the performance impact, we time the application run while the client system is also serving an intense write workload: writing 1GB files back to back.

The performance baseline for all results presented in the figures below is the execution time of the application on an unloaded system (i.e., neither the *MosaStore* SAI client nor other applications are running on the system at the time). We present averages over 20 runs.

**The impact on a compute bound application**. Figure 12, Figure 13, and Figure 14 present the *MosaStore* achieved throughput (left), and the compute bound application execution time increase (i.e., application slowdown – the lower, the better) (right) under the three workloads presented in Section 4.2. These results confirm that:

- First, outsourcing to the GPU frees CPU cycles that can be effectively used by a compute intensive competing application. In all three experiments, with different workloads, the competing application performs faster on a client system that offloads to the GPU than on a client system that computes hash functions on the CPU. The difference can vary from as high as reducing the slowdown by half (for the 'different' workload) to reducing the slowdown by 10-20% (for the other two workloads).
- Second, outsourcing to the GPU enables better storage system throughput (around 2.5x better under the 'similar' workload, and 2x better under the checkpoint workload) even when a competing compute intensive application is present on the client node.
- Third, the throughput of the GPU-enabled storage system is only slightly affected by the competing application: less than 18% throughput loss compared to the case when the client system runs on a dedicated node.

We note that the *non-CA* system imposes a significant burden on the competing application (between 225% and 80% slowdown depending on workload). This is due to high TCP processing overheads (we have verified this by running the competing application while continuously generating TCP traffic using *iperf*. In this case *iperf* caused 185% application slowdown).

A second, surprising, observation is that under the different workload, *CA-GPU* has lower impact on the competing compute-bound application than the *non-CA* system configuration (that does not consume CPU cycles for hashing). While we do not have a precise understanding for this performance disparity, our intuition is that it is related to the blocking of the SAI threads (introduced by GPU calls) that yield more frequently to the application.

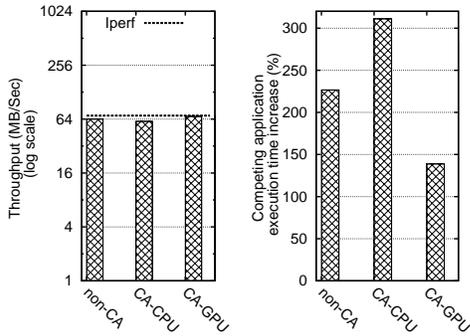

Figure 12: (Left) MosaStore average achieved throughput under the different workload while running a competing compute intensive application. (Right) Competing application slowdown (the lower the better).

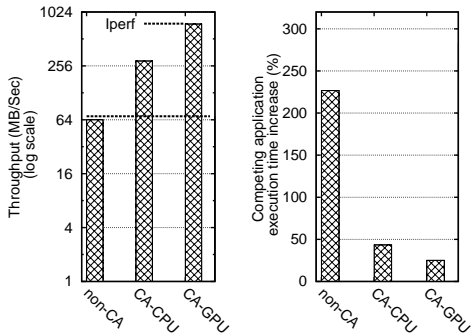

Figure 13: (Left) MosaStore average achieved throughput under the 'similar' workload while running a competing compute intensive application. (Right) Competing application slowdown (the lower, the better).

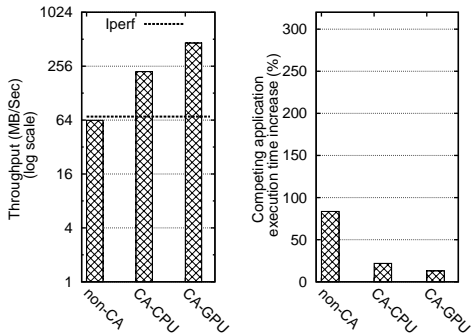

Figure 14: (Left) MosaStore average achieved throughput under the 'checkpoint' workload while running a competing compute intensive application. (Right) Competing application slowdown (the lower, the better).

**The Impact on an I/O bound application**. Figure 15, Figure 16, and Figure 17, present MosaStore's achieved throughput (left) and the disk IO-bound application slow down (right), under the three workloads presented in Section 4.2.

There are three main takeaways from these experiments:
- First, offloading to the GPU does not introduce a bottleneck for a competing I/O-intensive application. The competing application slowdown is marginally (5-15%) lower (thus better) than when hashing on CPU.
- Second, outsourcing to the GPU enables better storage system throughput (around 2x better under the 'similar' and 1.7x better under the checkpoint workload) even when a competing I/O intensive application is present on the client node.
- Third, the throughput of the GPU-enabled storage system is only slightly affected by the competing application: less than 6% throughput loss compared to the case when the client system runs on a dedicated node.

### 4.6 Summary

Our evaluation highlights five findings:
- First, our evaluation shows that offloading content addressability mechanism to GPU not only brings tangible performance gains; but, importantly, it opens the door for using content addressability in high performance computing (HPC) systems, an area where this mechanism that has been often avoided not only due to its raw computational overheads but mostly because it adds a performance bottleneck where high-speed networks had been deployed.
- Second, our evaluation shows that a task management engine (i.e., our CrystalGPU layer) that orchestrates GPU offloading, and provides application agnostic optimizations, is not only essential from a software architecture standpoint but also from a performance standpoint.
- Third, our evaluation shows that GPU offloading provides close to optimal performance.
- Fourth, GPU offloading does not introduce a new system bottlenecks: the impact on competing CPU-bound and IO-bound is negligible.
- Finally, our evaluation provides an important data point for system builders: for hashing-based workloads, extending the system with a GPU achieves significantly higher results than extending the system with a second CPU (of close market price).

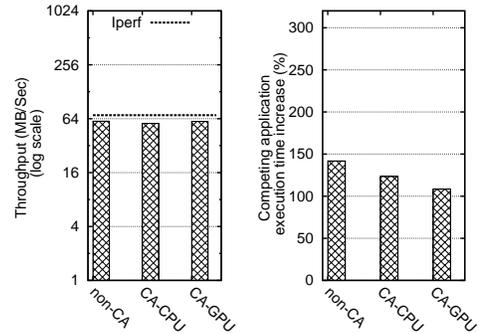

Figure 15: (Left) MosaStore's average achieved throughput while running an I/O intensive application with the different workload. (Right) Competing application slowdown (lower is better).

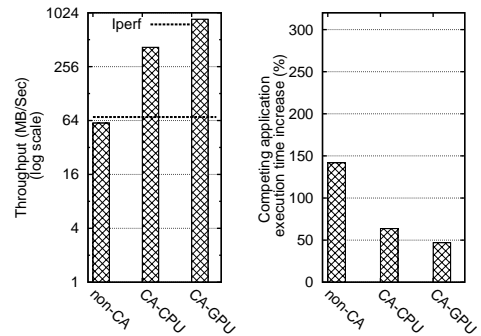

Figure 16: (Left) MosaStore's average achieved throughput while running an I/O intensive application with the similar workload. (Right) Competing application slowdown (lower is better).

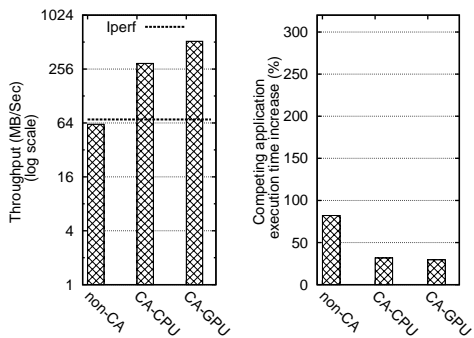

Figure 17: (Left) MosaStore's average achieved throughput while running an I/O intensive application with the checkpoint workload. (Right) Competing application slowdown (lower is better).

## 5 RELATED WORK

The background and the introduction have included positioned this project relative to important aspects of related work. This section adds two new dimensions.

*Hashing.* Hashing is commonly used by storage systems to support: content addressability [1, 2], data integrity checking [27, 30, 31], load balancing [32, 33], data similarity detection [3, 21], deduplication backup operations [12], and version control (e.g, rsync [34]). The use of hashing in these systems differs along multiple axes. We enumerate only two: block sizes - from fixed-sized blocks [1, 2] that use hashing similar to our direct-hashing module, to detecting block boundaries based on content [3, 12, 21] and use, what we call sliding-window hashing in Section 2.1; and targeted deployment environment - from personal use [2], peer to peer [30] to data center [12]. Other uses of hashing include: detection of copyright violations [35, 36], compact set representation [37], and various security mechanisms [27].

*GPU harnessing.* Exploiting GPUs for general purpose computing has recently gained popularity, particularly as a mean to increase the performance of scientific applications. We refer the reader to Owens et al. [38] for a comprehensive survey.

More related to our infrastructural focus, Curry et al. [19] explore the feasibility of using GPUs to enhance RAID system reliability. Their preliminary results show that GPUs can be used to accelerate Reed Solomon codes [39], used in RAID6 systems. Along the same lines Falcao et al. [40] show that GPUs can be used to accelerate Low Density Parity Checks (LDPC) error correcting codes.

Harrison and Waldron [41], study the feasibility of GPUs as a cryptographic accelerator. To this end they implement the AES encryption algorithm using CUDA and report 4x speedup. Finally, Moss et al. [42] study the feasibility of accelerating the mathematical primitives used in RSA encryption. They use OpenGL and render the application into a graphics application and report up to 3x speedup. Moreover, recently exploiting GPUs to accelerate security operations was adopted in few security products, including Kaspersky antivirus [43] and Elcomsoft password recovery software [44].

Our study is different from the above studies in two ways. First, unlike the previous studies that focus on accelerating standalone primitives, we evaluate the viability of exploiting the GPU computational power in the context of a complete system design. Second, the hashing primitives we focus on are data-intensive with a ratio of computation to input data size of at least one order of magnitude lower than in previous studies.

Finally, the storage systems research community has, recently, adopted the design approach proposed by this stream of work: to exploit the GPUs as storage system accelerators. Two recent studies adopt this approach: PTask project [4] proposes an OS abstraction for GPU management and demonstrates the PTask framework efficiency through building an encrypted storage system that offloads the encryption computation to GPUs. Further, Shredder [5] project adopts the design and the optimization techniques proposed in this work to build a deduplicated storage system for MapReduce incremental computation system.

## 6 CONCLUSIONS

At a high level of abstraction, computing system design is a multi-dimensional constraint optimization problem: designers have at hand various components that offer different cost-to-performance characteristics as well as the techniques to put these components together. These techniques have their own overheads that lead to performance tradeoffs and new limitations. In this context, a system designer optimizes for key success metrics (such as latency, throughput, storage footprint, and data durability) within specific constraints (such as cost and availability).

Historically, the unit costs for raw storage, computation, or network transfers have evolved largely in sync. Recently, however, massively multi core commodity hardware (such as GPUs) holds the promise of a sharp, one order of magnitude, drop in computation costs. This drop in the cost of computation, as any order-of-magnitude drop in the cost per unit of performance for a class of system components, triggers the opportunity to redesign systems and to explore new ways to engineer them to recalibrate the cost-to-performance relation.

This study demonstrates the feasibility of exploiting GPU computational power to support distributed storage systems, by accelerating popular compute-intensive primitives. We prototyped a storage system capable of harvesting GPU power, and evaluated it under two system configurations: as a content addressable storage system and as a traditional system that uses hashing to preserve data integrity. Further, we evaluated the impact of offloading to the GPU on competing applications' performance, and the impact of competing applications on the storage system performance. Our results show that this technique can bring tangible performance gains enabling close to optimal system performance under the set of workloads and configurations we considered without negatively impacting the performance of concurrently running applications.

Whereas we have effectively demonstrated the feasibility of tapping into GPUs computational power to accelerate compute-intensive storage system primitives, we view this effort as a first step towards the larger goal of understanding how we can use heterogeneous multicore processors for independent system-level enhancements rather than application-level speedups.

We note that future single-chip massively multicore processors will likely be similar to our experimental platform along two directions: they will be heterogeneous (e.g., include different cores with different execution models, SIMD vs. MIMD) and will offer complex application manageable memory hierarchies. In this context performance gains can only be obtained by careful orchestration of data transfers, data placement, and task allocation. Our experience with a combination of parallel hashing on a GPU, optimized memory handling, and task scheduling leads us to believe that these goals can be effectively be mapped to independent abstraction layers.

Processor manufacturers expect to develop many-core architectures over the next decade [16]. While we have proposed a solution that addresses germane problems for storage systems with current technology, the high-level principles embodied in our work will extend to the many-core arena as well. Some of the challenges that we encountered in our development relate to the efficient handling of memory transfers and shared memory. Such concerns are likely to exist even with many cores on a single chip; it is expected that many core architectures will utilize heterogeneous cores and non-uniform memory accesses (NUMA) and such architectures will continue to require careful memory management to achieve significant speedups.

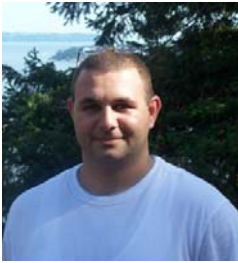
**Samer Al-Kiswany** received the MSc degree from the University of British Columbia (UBC), where he is currently working toward the PhD degree in the Electrical and Computer Engineering Department. His research interests are in distributed systems with special focus on high performance computing systems, and cloud computing. He is a student member of the IEEE.

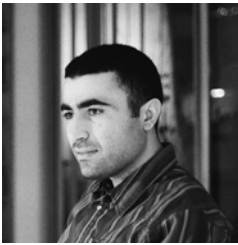
**Abdullah Gharaibeh** received the MASc degree in computer engineering from the University of British Columbia. He is currently working toward the PhD degree in the Computer Engineering Department at the University of British Columbia. His interests are in the design and evaluation of high-performance distributed systems and GPU-based computing.

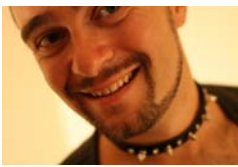
**Matei Ripeanu** received the PhD degree in computer science from The University of Chicago. He is currently an associate professor with the Computer Engineering Department of the University of British Columbia. Matei is broadly interested in large-scale distributed systems.